\documentclass[aps,reprint,twocolumn]{revtex4-1}
\usepackage[english]{babel}
\usepackage{times}
\usepackage[cp1250]{inputenc}
\usepackage{graphicx, texdraw}
\usepackage{color}
\usepackage{overpic} %pakiet do umieszczania napisów na rysunkach!
\usepackage{amsmath,amsthm,latexsym,amssymb,amsfonts,epsfig}
\newcommand{\be}{\begin{equation}}
\newcommand{\ee}{\end{equation}}
\newcommand{\ba}{\begin{eqnarray}}
\newcommand{\ea}{\end{eqnarray}}

\usepackage[mathscr]{euscript} %Euler script

\newcommand{\br}{\mathbf{r}}
\newcommand{\bv}{\mathbf{v}}
\newcommand{\bk}{\mathbf{k}}

\newcommand{\de}{\mathrm{d}}

\usepackage[colorlinks=true, citecolor=blue,  ]{hyperref}
\begin{document}
\title{Four approaches to hydrodynamic Green's functions - the Oseen tensors}
\author{Maciej Lisicki}
\email{Electronic address: mklis@fuw.edu.pl}
\address{Institute of Theoretical Physics, Faculty of Physics, University of Warsaw  \\ ul. Ho\.za 69, 00-681 Warszawa, Poland}
\begin{abstract}
We present four different ways of deriving the Oseen tensor which is the fundamental solution to the Stokes equations for an incompressible viscous fluid. This solution corresponds to a point force acting on an infinite fluid. The derivations follow the books of Kim \& Karilla, Zapryanov \& Tabakova, Dhont, and Pozrikidis.
\end{abstract}
\maketitle

%==============================================================================
{\it The shortest path between two truths in the real domain passes through
the complex domain.}
{\begin{flushright}Jacques Hadamard\end{flushright}
\vspace{5pt}

In this short article, we present several different ways of deriving the Green's functions for Stokes equations for an infinite viscous fluid. The fundamental solution for velocity is known as the Oseen tensor, while the corresponding solution for pressure is simply referred to as the pressure vector. 

The derivations follow four well-known books. The solutions base either on the properties of Fourier transformation \cite{Kim,Zapr}, the linearity of Stokes equations \cite{Dhont}, or the properties of fundamental solutions of Laplace and biharmonic equations \cite{Pozrikidis}.

\section{Formulation of the problem}
We shall focus on the following problem: a force $\mathbf{F}$ is applied to a point particle immersed in space filled with viscous liquid. We are looking for the fundamental solution (\textit{stokeslet}) in case of incompressible flow. The full system of flow equations in the low Reynolds number regime ($\mathrm{Re}\ll1$) has the form

\ba -\nabla p(\br)+\mu\nabla^2 \bv(\br) &=& -\mathbf{F}\delta(\br), \label{Stokes} \\ 
\nonumber 
\nabla\cdot\bv &=& 0. \ea
We demand that the flow field vanishes at infinity, i.e. $|\bv(\br)|\stackrel{r\to\infty}{\longrightarrow}0$. Then the solution is uniquely determined.
By solving this problem we mean finding the form of velocity and pressure fields in the fluid.

%===========================================================================
%===========================================================================
%===========================================================================
\section{Solution \`a la Kim \& Karilla}
A very 'physical' approach to this phenomenon can be made by making use of the linearity of Stokes equations and symmetries of the system. Here, we preform detailed calculations according to hints from Exercise \textbf{2.9} from the textbook \cite{Kim}.

As the Stokes equations are linear in $p$ and $\bv$, the pressure field $p(\br)$ can be written as a scalar product of a certain vector field $\mathbf{P}(\br)$ and the point force $\mathbf{F}$ and the velocity field $\bv(\br)$ can be represented by a  a tensor $\mathbb{G}(\br)$ acting on a point force $\mathbf{F}$ with certain factors:
\be p(\br)=\frac{\mathbf{F}\cdot \mathbf{P}(\br)}{8\pi\mu}, \qquad \bv(\br)=\frac{\mathbb{G}(\br)\cdot\mathbf{F}}{8\pi\mu}. \ee
Applying Einstein's summation rule we can write these equations using the vector components 
$$ p(\br)=\frac{P_j F_j}{8\pi\mu}, \qquad v_i(\br)=\frac{\mathbb{G}_{ij} F_j}{8\pi\mu}. $$

Applying the Fourier transform (\hyperref[Fourier]{App. A.1}) to Stokes equations (\ref{Stokes}), we get
$$ -i\bk \widehat{p}-\mu k^2 \widehat{\bv} = -\mathbf{F}. $$
Inserting the expected field form to this equation we have for component $i$:
$$ -ik_i\frac{\widehat{P}_j }{8\pi\mu}F_j- k^2 \frac{\widehat{\mathbb{G}}_{ij}}{8\pi} F_j = - F_i = -F_j\delta^K_{ij}. $$
We have introduced the Kronecker delta to eliminate $F_j$ and therefore can write
\be -ik_i\frac{\widehat{P}_j}{8\pi\mu} - k^2 \frac{\widehat{\mathbb{G}}_{ij}}{8\pi} = -\delta^K_{ij}. \label{StoFour}\ee
The incompressibility condition reads
$$ k_i\mathbb{G}_{ij}=0. $$
Multiplying the Stokes equation in the form as above and using the incompressibility relation we eliminate the velocity tensor part of the equation and get
$$ \frac{\widehat{P}_j}{8\pi\mu}=-\frac{i k_j}{k^2} $$
We can directly calculate the inverse Fourier transform of this equation to get $P_j$
\be  \frac{P_j}{8\pi\mu}= -\frac{i}{(2\pi)^3}\int_{\mathbb{R}^3} \de \bk \frac{k_j}{k^2}e^{i\bk\cdot\br} \label{pejot}\ee
Let us now consider a function $\phi=\frac{1}{4\pi r}$. One can show that its Fourier transform equals $k^{-2}$. Indeed, let us calculate the inverse transform. Choosing $\br$ to be parallel to the $x_3$ axis in the $\bk$-space, we can perform integration in spherical coordinates (fig. \ref{fig:int})
\ba \mathcal{F}^{-1}\{k^{-2}\}&=& \frac{1}{(2\pi)^2}\int_{0}^{\infty} \de k k^2 \int\limits_{-1}^{1}\frac{e^{ikr\cos\theta}}{k^2} \de(\cos\theta) \\ \nonumber
&=& \frac{2}{(2\pi)^2r} \int_{0}^{\infty} \de (kr) \frac{\sin{kr}}{kr} =\frac{1}{4\pi r}, \ea
\begin{figure}
	\centering
		\includegraphics[width=4cm]{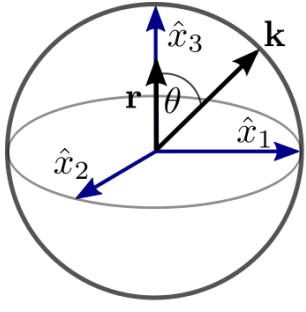}
		\caption{Integration in the $\bk$--space with fixed orientation of $\br=r \cdot\hat{x}_3$.}
		\label{fig:int}
\end{figure}
where we have used the (\ref{aa1}) integral (see \hyperref[Integral]{App. A.2}). By taking the gradient of $\phi$, we reproduce the Fourier terms from eq. (\ref{pejot}). We can therefore explicitly write down the solution for pressure:
\be \mathbf{P}(\br)=-2\mu \nabla\Bigl(\frac{1}{r}\Bigr). \label{pj} \ee
Now we shall use expression for $\widehat{P_j}$ to construct the velocity field. We insert $\widehat{P_j}$ into eq. (\ref{StoFour}) and get a closed expression for the $\mathbb{G}$ tensor
\be \frac{\widehat{\mathbb{G}}_{ij}}{8\pi} =\frac{\delta^K_{ij}}{k^2}-\frac{k_i k_j}{k^4} . \ee
Once again, the first term on the RHS is the transform of the Green's function for Laplace equation and after inversion contributes 
$\frac{\delta^k_{ij}}{4\pi r}$. It remains to find the inverse transform
\be \mathcal{F}^{-1}\Bigl\{\frac{k_i k_j}{k^4}  \Bigr\} \label{fodwr} \ee
Now we can use the symmetry argument to deduce the form of the inverted transform. We know that in the $\bk$-space $\widehat{\mathbb{G}}$ decreases as $k^{-2}$. This condition transforms to the real space resulting in an asymptotic decay of $\mathbb{G}$ as $r^{-1}$. Moreover, the expression (\ref{fodwr}) is symmetric with respect to index switching. Therefore we deduce the form of the inverse transform to be
$$ \mathcal{F}^{-1}\Bigl\{\frac{k_i k_j}{k^4}  \Bigr\}= C_1\frac{\delta^K_{ij}}{8\pi r} +C_2\frac{x_i x_j}{8\pi r^3}$$
The last task is to determine the values of numerical constants $C_1$, $C_2$. Taking the trace of both sides, we get one condition $3C_1+C_2=2$. Taking now $i=j=3$ and again fixing $\mathbf{r}=(0,0,x_3)$ in Cartesian coordinates, we perform the integration in $\bk$--space. 
\ba \nonumber & &\mathcal{F}^{-1}\Bigl\{\frac{k_i k_j}{k^4}  \Bigr\}=\frac{1}{(2\pi)^3} \int\limits_\mathbb{R} \de \bk \frac{k_3^2}{k^4} e^{i\bk\cdot\br}=\left[k_3=k\cos\theta \right]= \\ \nonumber
& & \frac{1}{(2\pi)^2} \int\limits_0^\infty \de k \int\limits_{-1}^1 \de(\cos\theta) \cos^2\theta e^{ikr\cos\theta}= \\ \nonumber
& & \frac{1}{(2\pi)^2} \int\limits_0^\infty \de k \int\limits_{-1}^1 \de\xi\xi^2 e^{ikr\xi} = (\ast) \ea
Performing integration by parts, we transfer the $\xi$--integral into known definite integrals (\ref{aa1},\ref{a2},\ref{a3}).
\ba \nonumber 
 &(\ast)&=\frac{1}{(2\pi)^2} \int\limits_0^\infty \de k \Bigl[\frac{2\sin kr}{kr}-\frac{2}{ikr} \int\limits_{-1}^1 \de\xi\xi e^{ikr\xi} \Bigr]= \\ \nonumber
&=&\frac{1}{(2\pi)^2} \int\limits_0^\infty \de k \Bigl[\frac{2\sin kr}{kr}-\frac{4\cos kr}{(kr)^2} -\frac{2}{(kr)^2}\int\limits_{-1}^1 \de\xi\ e^{ikr\xi} \Bigr] \\ \nonumber
&=& \frac{1}{(2\pi)^2 r} \int\limits_0^\infty \de u \Bigl[\frac{2\sin u}{u}-\frac{4\cos u}{u^2} -\frac{4\sin u}{u^3} \Bigr]=0 \ea
We therefore must have $C_1+C_2=0$. From these two conditions we can deduce the constants $C_1=-C_2=1$ and sum up the results writing 
\be {\mathbb{G}}_{ij} =\frac{\delta^K_{ij}}{r}+ \frac{r_i r_j}{r^3}, \qquad \mathrm{or} \qquad \mathbb{G} = 
\frac{1}{r}\Bigl(\mathbb{I}+\frac{\br\otimes\br}{r^2}\Bigr). \ee
The pressure and velocity fields can be written as:
\be  \boxed{ p(\br)= \frac{\mathbf{F}\cdot\br}{4\pi r^3}  \qquad \bv(\br)=\frac{\mathbf{F}}{8\pi\mu r}\cdot\Bigl(\mathbb{I}+\frac{\br\br}{r^2}\Bigr)}\ee
%

%==============================================================================
%==============================================================================
%==============================================================================
%
\section{Solution \`a la Zapryanov \& Tabakova}
This method has been proposed by \cite{Zapr}. It follows from the properties of Fourier transform and fundamental solutions of Laplace and biharmonic equations. In fact, one does not have to calculate any integral explicitly in this approach.

We start by taking the divergence of eq. (\ref{Stokes}), obtaining a Poisson equation for pressure
$$ \nabla^2 p(\br)=\nabla(\mathbf{F}\delta(\br)). \label{Poisson}$$
We can now apply Fourier transform to the above equation, which gives 
$$ k^2 \widehat{p}(\bk) = -i\bk\cdot \mathbf{F} \quad \Longrightarrow  \widehat{p}=-i\frac{\bk\cdot\mathbf{F}}{k^2}$$
We can now apply the Fourier transform to Stokes equations and use the obtained expression to eliminate pressure from the equations and get a closed expression for the velocity field:
$$  i\bk \Bigl(\frac{\bk\cdot\mathbf{F}}{k^2}\Bigr) + \mu k^2 \widehat{\bv}(\bk) = \mathbf{F}. $$
We find the expression for the velocity field in $\bk$-space:
\be \widehat{\bv}(\bk)=  \frac{1}{\mu k^2}\Bigl[\mathbf{F} - \bk \Bigl(\frac{\bk\cdot\mathbf{F}}{k^2}\Bigr)\Bigr].\ee
Using the definition of the inverse Fourier transform, we can easily write down the expressions for pressure and velocity fields
\ba \bv(\br) &=& \frac{1}{8 \mu \pi^3}\int\limits_{\mathbb{R}^3} \de \bk \frac{e^{i\bk\cdot\br}}{k^2} 
\Bigl[\mathbf{F} -  \bk \Bigl(\frac{\bk\cdot\mathbf{F}}{k^2}\Bigr)\Bigr] \label{velo} \\ 
p(\br) &=& \frac{i}{8\pi^3}\int\limits_{\mathbb{R}^3} \de \bk  \frac{e^{i\bk\cdot\br}}{k^2}(\bk\cdot\mathbf{F}) \label{press}. \ea
We can proceed to solution of these equations. Basing on the knowledge of fundamental solutions of the Laplace equation (see \hyperref[Fourier]{App. A.1}), we find useful expressions relating derivatives of $1/r$ with their Fourier transforms. We now note that by taking a scalar product of  eq. (\ref{2312}) with  $\mathbf{F}$, we get an integral which describes the pressure field. Inserting it into the relation (\ref{press}), we get
\be \boxed{ p(\br)=-\mathbf{F}\cdot\nabla\Bigl(\frac{1}{4\pi r}\Bigr)=\frac{\mathbf{F}\cdot\br}{4\pi r^3}} \ee
In a similar manner, taking the scalar product of eq. (\ref{2313}) with $\mathbf{F}$, we reproduce the integral present in the first part of the expression for velocity field (\ref{velo}), so that
\be \bv(\br)=\frac{\mathbf{F}}{4\pi\mu r} - \frac{1}{8 \mu \pi^3}\int\limits_{\mathbb{R}^3} \de \bk \frac{e^{i\bk\cdot\br}}{k^2}  \bk \Bigl(\frac{\bk\cdot\mathbf{F}}{k^2}\Bigr) \ee
Taking a scalar product of eq. (\ref{2320}) with $\mathbf{F}$, we get the second term of the RHS of velocity field equation. We have used the identity $\bk\bk \cdot \mathbf{F} = (\mathbf{F}\cdot \bk)\bk$. These relations imply
\be \bv(\br)=\frac{\mathbf{F}}{4\pi\mu r} -\frac{\mathbf{F}}{\mu} \cdot \nabla\otimes\nabla\Bigl(\frac{r}{8\pi }\Bigr) \label{tensorki} \ee
Due to the fact that $\nabla r=\frac{\br}{r}$ and $\nabla\otimes\br=\mathbb{I}$ - the unit tensor, we can write
$$ \frac{1}{8\pi}\nabla\otimes(\nabla r))  = \frac{1}{8\pi}\nabla\otimes\Bigl(\frac{\br}{r}\Bigr)=-\frac{1}{8\pi}\frac{\br\br}{r^3}+\frac{\mathbb{I}}{8\pi r}, $$
where we have used the Leibniz chain rule for calculation of $\nabla\otimes\frac{\br}{r}$. We can always write a vector in a form $\mathbf{F}=\mathbb{I}\mathbf{F}$, so that in the end we can write down the expression for velocity field
\be \boxed{\bv(\br)=\frac{\mathbf{F}}{8\pi\mu r}\cdot\Bigl(\mathbb{I}+\frac{\br\br}{r^2}\Bigr)}\ee

%==============================================================================
%==============================================================================
%==============================================================================

\section{Solution \`a la Dhont}
A very physical approach, basing on the linearity of Stokes equations has been presented by \cite{Dhont}. In this approach we clearly see the Green's functions role in the solution of the problem of external force density in the fluid and illustration of the superposition principle.

Consider an external force $\mathbf{F}$ acting on a fluid only in a single point $\br'$, so that $\mathbf{F}(\br)=\mathbf{F}\delta(\br-\br')$. Since the Stokes equations (\ref{Stokes}) are linear, the fluid flow velocity in a point $\br$ somewhere in the fluid is proportional to the force and has to depend on the direction of the force and the distance to the point where the force is exerted. Moreover, this relation has to be a linear transformation (which can be represented as a matrix). Hence, in a natural way, we can write
\be \bv(\br)=\mathbb{T}(\br-\br')\cdot\mathbf{F}. \label{VelDh}\ee
Similarly, pressure is linearly related to the force by a vector quantity
\be p(\br)=\mathbf{g}(\br-\br')\cdot\mathbf{F}. \label{PressDh}\ee
The usual terminology is the \textit{Oseen tensor} for $\mathbb{T}$ and the \textit{pressure vector} for $\mathbf{g}$.

Consider now an external force which is continuously distributed over the entire fluid (i.e. there is a nonzero external force density $\mathbf{f}(\br')$ in the fluid). The linearity of Stokes equations implies the superposition principle - the fluid velocity in a certain point $\br$ is a vector sum of the fluid velocity increments stemming from the forces acting in every point of the fluid. We can therefore express this sum as an integral
$$ \bv(\br)=\int\de \br' \mathbb{T}(\br-\br')\cdot\mathbf{f}(\br'). $$
For pressure the same arguments hold and we can write
$$ p(\br)=\int\de \br' \mathbf{g}(\br-\br')\cdot\mathbf{f}(\br'). $$
In this structure one immediately sees the role of Green's functions for linear problems. Once the force field is specified, knowing the Green's functions for a particular geometry, the pressure and velocity are easily found by integration. Deriving the Green's functions needs an 'inversed' reasoning. It involves solving the Stokes equations with a specific force field, namely a point force (concentrated in one point, what is represented by a Dirac delta function).
Let us substitute eq. (\ref{VelDh}) and (\ref{PressDh}) into the Stokes equations . We get
\ba & &\int\de \br' [\nabla \cdot\mathbb{T}(\br-\br')]\cdot\mathbf{f}(\br'), = \mathbf{0} \\ \nonumber
& &\int\de\br' [\nabla\mathbf{g}(\br-\br')-\mu \nabla^2 \cdot\mathbb{T}(\br-\br')-\mathbb{I}\delta(\br-\br')]\cdot \mathbf{f}(\br') = \mathbf{0},
\ea
where we have used an identity $\mathbf{F}=\int\de\br' \mathbf{f}(\br')\delta(\br-\br')$. Since the external force density is arbitrary, the expressions in brackets must vanish and hence we get the equations for the Green's functions
\ba \nabla \cdot\mathbb{T}(\br-\br') &=& 0, \\ 
\nabla\mathbf{g}(\br-\br')-\mu \nabla^2 \cdot\mathbb{T}(\br-\br')-\mathbb{I}\delta(\br-\br') &=& \mathbf{0}. \label{tens}
\ea
In this equation we have now tensor quantities and we shall in fact write $\nabla\otimes\mathbf{g}$ instead of $\nabla \mathbf{g}$.
An usual operation is now to take the divergence of the second equation and use the incompressibility condition to get a Poisson equation fo the pressure vector:
$$ \nabla^2 \mathbf{g} = -\nabla\cdot\mathbb{I}\delta(\br)=\delta(\br) $$
We can now recall the fundamental solution of the Laplace equation (\ref{fund}), which yields the form of $\mathbf{g}$
$$ \mathbf{g}=-\nabla\Bigl(\frac{1}{4\pi r}\Bigr) +\mathbf{G}(\br), $$
as we can alway add a vector $\mathbf{G}(\br)$ which satisfies the laplace equation, i.e. $\nabla^2\mathbf{G}=0$. One can show that if we demand an asymptotic decay so that $\mathbf{G}\to \mathbf{0}$ as $r\to\infty$, this implies $\mathbf{G}\equiv \mathbf{0}$ (see \hyperref[Laplace]{App. A.4}). Hence, we find the pressure vector
\be \boxed{\mathbf{g}=-\nabla\Bigl(\frac{1}{4\pi r}\Bigr)=\frac{1}{4\pi}\frac{\br}{r^3}}  \label{pressfield}\ee
Now we can substitute (\ref{pressfield}) to eq. (\ref{tens}). Noting that $\nabla\otimes\frac{\br}{r^3}= \frac{\mathbb{I}}{r^3}-3\frac{\br\otimes\br}{r^5}$ and eliminating the Dirac delta by substituting the fundamental solution of Laplace equation, one gets
\be \nabla^2\Bigl[\frac{1}{4\pi}\frac{\mathbb{I}}{r}-\mu\mathbb{T}(\br)\Bigr]=\frac{1}{4\pi}\Bigl[3\frac{\br\br}{r^5}-\frac{\mathbb{I}}{r^3}\Bigr]\ee 
We now choose, basing on the RHS form of the above equation, the appropriate form of the LHS to be
\be \nabla^2\Bigl[\frac{1}{4\pi}\frac{\mathbb{I}}{r}-\mu\mathbb{T}(\br)\Bigr]=\alpha_0\frac{1}{r^n}\mathbb{I}-\alpha_1\frac{1}{r^m}\frac{\br\br}{r^2}, \ee
where $\alpha_0,\ \alpha_1,\ m$ and $n$ are constants. These constants can be chosen in such a way that this Ansatz provides a solution for the Oseen tensor decaying at inifinity ($\mathbb{T}(\br)\to 0$ as $t\to\infty$) and after some algebra we arrive at the known result
\be \boxed{\mathbb{T}(\br)=\frac{1}{8\pi\mu r}\cdot\Bigl(\mathbb{I}+\frac{\br\br}{r^2}\Bigr).}\ee

%==============================================================================
%==============================================================================
%==============================================================================
\section{Solution \`a la Pozrikidis}
The fourth method in our review was proposed by \cite{Pozrikidis}. In this case we consider Stokes equations with arbitrary force $\mathbf{F}$, concentrated on one point. Because pressure is a harmonic function, and replacing the delta function on the RHS of eq. (\ref{Stokes}) basing on the fundamental solution of Laplace equation, we can set (to balance the dimensions of pressure)
$$p=-\frac{1}{4\pi}\mathbf{g}\cdot\nabla\Bigl(\frac{1}{r}\Bigr).$$
This can be obtained by taking the divergence of Stokes equation and then replacing the delta function with $-\nabla(\frac{1}{4\pi r})$.
Substituting the delta function and the pressure function to the Stokes equations, we get
\be \mu\nabla^2 \bv =-\frac{1}{4\pi}\mathbf{g}(\nabla\nabla - \mathbb{I}\nabla^2)\Bigl(\frac{1}{r}\Bigr). \label{cos}\ee
We can introduce a scalar function $H$ and express the velocity field in the form
\be \bv=\frac{1}{\mu} \mathbf{g}\cdot (\nabla\nabla - \mathbb{I}\nabla^2) H. \label{cos2}\ee
One can prove that such an operation can always be preformed and $H$ can be found.
By replacing $\bv$ in eq. (\ref{cos}), we arrive at a closed expression for $H$ (if we discard an arbitrary constant $\mathbf{g}$):
$$(\nabla\nabla - \mathbb{I}\nabla^2) \Biggl( \nabla^2 H +\frac{1}{4\pi r}\Biggr)=0. $$
This equality can be surely satisfied by any solution of the Poisson equation $\nabla^2 H=-\frac{1}{4\pi r}$. We therefore find (by applying the Laplace operator to this condition) that $H$ satisfies the biharmonic equation $\nabla^4 H = \delta(\br)$. We know the form of the fundamental solution ((see \hyperref[fundam]{App. A.3}) so that
$$ H=-\frac{r}{8\pi} $$
Substituting this result into eq. (\ref{cos2}), we arrive (after some algebra, the same as in eq. (\ref{tensorki}) at our result
$$v_i(\br)=\frac{1}{8\pi\mu}\mathbb{S}_{ij} g_j,$$
where the Oseen tensor $\mathbb{S}_{ij}$ is defined as follows:
$$\mathbb{S}_{ij}(\br)= \frac{\delta_{ij}}{r}+\frac{x_i x_j}{r^3} $$

\section*{Acknowledgements}
The Author wishes to thank Maria Ekiel-Je\.zewska for the idea of this report and discussions.
%==============================================================================
%==============================================================================
%==============================================================================

%
%
% DODATEK
%
%==============================================================================
\appendix
\section{Mathematical addendum}
\subsection{Fourier transform \label{Fourier}}
We define the Fourier transform pair in the following manner
\ba \nonumber \mathcal{F}\{f\}=\widehat{f}(\bk) &=& \int_{\mathbb{R}^3} \de \br f(\br)e^{-i\bk\cdot\br}, \\ \nonumber
\mathcal{F}^{-1}\{f\}=f(\br) &=& \frac{1}{(2\pi)^3}\int_{\mathbb{R}^3} \de \bk \widehat{f}(\bk)e^{i\bk\cdot\br}. \ea
\subsection{Useful integrals \label{Integral}}
We have the following integrals
\ba  \int\limits_0^\infty \de k \frac{\sin k}{k} &=&\frac{\pi}{2} \label{aa1} \\
\int\limits_0^\infty \de k \frac{1-\cos k}{k^2} &=&\frac{\pi}{2} \label{a2}\\
 \int\limits_0^\infty \de k \frac{k-\sin k}{k^3} &=&\frac{\pi}{4} \label{a3} \ea
The second and third integral actually can be obtained from integration by parts of the first integral.
%==============================================================================

\subsection{Fundamental solutions \label{fundam}}
We introduce fundamental solutions for Laplace and biharmonic equations. These are such functions $\psi$ and $\phi$ that satisfy the corresponding equations with Dirac delta functions:
\ba \nonumber \nabla^2 \phi(\br) &=& -\delta(\br) \\ \nonumber
\nabla^4 \psi(\br) &=& -\delta(\br) \ea
It appears that in 3D the fundamental solutions have the form
\be \phi=\frac{1}{4\pi r}; \quad \psi=\frac{r}{8\pi} \label{fund} \ee
Finding these solutions involves inverting Fourier transforms of the corresponding equations and involves integration in complex plane. 

By differentiating the fundamental solution of Laplace equation we get the corresponding terms in Fourier space
\ba \nabla^2 \Bigl(\frac{1}{4\pi r}\Bigr) &=& -\delta(\br) = - \frac{1}{(2\pi)^3}\int_{\mathbb{R}^3} \de\bk e^{i\bk\cdot\br}, \\
\nabla \Bigl(\frac{1}{4\pi r}\Bigr) &=&   -\frac{i}{(2\pi)^3}\int_{\mathbb{R}^3} \de \bk \frac{\bk}{k^2} e^{i\bk\cdot\br}, \label{2312} \\
\Bigl(\frac{1}{4\pi r}\Bigr) &=&  \frac{1}{(2\pi)^3}\int_{\mathbb{R}^3} \de \bk \frac{1}{k^2} e^{i\bk\cdot\br}. \label{2313} \ea
Analogical operations on fundamental solutions of biharmonic equation lead to
\ba
\nabla^4 \Bigl(\frac{r}{8\pi}\Bigr) &=& -\delta(\br) = - \frac{1}{(2\pi)^3}\int_{\mathbb{R}^3}\de \bk e^{i\bk\cdot\br}, \\
\Bigl(\frac{r}{8\pi }\Bigr) &=&  - \frac{1}{(2\pi)^3}\int_{\mathbb{R}^3} \de \bk \frac{1}{k^4} e^{i\bk\cdot\br}, \\
\nabla\nabla\Bigl(\frac{r}{8\pi }\Bigr) &=&  \frac{1}{(2\pi)^3}\int_{\mathbb{R}^3} \de \bk \frac{\bk \bk}{k^4} e^{i\bk\cdot\br}. \label{2320} \ea
The last term contains tensor product of two vectors: $\nabla\nabla(\frac{r}{8\pi }) \equiv \nabla \otimes[\nabla(\frac{r}{8\pi })]$ and $\bk\bk \equiv \bk\otimes\bk$. This notation defines a second--rank tensor which can be represented as a matrix. In terms of matrix components one can write $(\bk\bk)_{ij} = k_i k_j$. We have created a tensor on the LHS of the last equation so that the character of the RHS is the same in the $\bk$--space.
%==============================================================================

\subsection{Laplace equation solutions' properties \label{Laplace}}
We encounter the problem of showing that a function $\mathbf{G}(\br)$ satisfying the following conditions:
\ba \nonumber \nabla^2f(\br) = 0\ & &\mathrm{on}\ \mathbb{R}^3, \\ \nonumber
f(\br) \to 0\ & &\mathrm{for}\ r\to\infty, \ea
is identically equal to 0. 

For this, we use the Green's integral formula for two scalar fields $\phi,\psi$:
$$\int_V \de \br (\phi\nabla^2\psi)=\int_{\partial V}\phi (\mathbf{n}\cdot(\nabla\psi) \de S-\int_V \de \br (\psi\nabla^2\phi) $$
Taking $\psi =f(\br')$ and $\phi=\frac{1}{|\br-\br'|}$, the boundary integral vanishes as $r\to\infty$ and we get
$$ \int \de \br' f(\br') \nabla'^2 \frac{1}{|\br-\br'|}=- \int \de \br' \frac{1}{|\br-\br'|} \nabla'^2 f(\br') = 0,$$
as $f$ satisfies the Laplace equation. But we know the fundamental solution of the Laplace equation, so that $\nabla'^2 \frac{1}{|\br-\br'|}=-4\pi\delta(\br-\br')$. From the above equation we deduce
$$4\pi f(\br)= 0 \Longrightarrow\ f(\br)\equiv 0.$$
We have proved that a harmonic function decreasing to 0 at infinity is identically 0 everywhere.
\end{document}